\documentclass[aps,twocolumn]{revtex4}
\usepackage[dvips]{graphics,graphicx}
\usepackage{amsmath}
\begin{document}

\title{Mott-insulator state of cold atoms in tilted optical lattices:  doublon dynamics and multi-level Landau-Zener tunneling}
\author{Andrey~R.~Kolovsky$^{1,2}$}
\author{Dmitrii N. Maksimov$^{1}$}
\affiliation{$^1$Kirensky Institute of Physics, 660036 Krasnoyarsk, Russia}
\affiliation{$^2$Siberian Federal University, 660041 Krasnoyarsk, Russia}
\date{\today}
\begin{abstract}
We discuss the dynamical response of strongly interacting Bose atoms in an adiabatically tilted optical lattice. The analysis is performed in terms of the multi-level Landau-Zenner tunneling. Different regimes of tunneling are identified and analytical expressions  for the doublon number, which is the quantity measured in laboratory experiments, are derived.
\end{abstract}
\maketitle

%%%%%%%%%%%%%%%%%%%%%%%%%%%%%%%%%%%%%%%%%%%%%%%%%%
\section{Introduction}

Experimental demonstration of the Mott insulator (MI) state with cold atoms in 2002 \cite{Grei02}  sparkled the interest to the controlled excitation of highly correlated many-body states. One of possible techniques to achieve such an excitation is application of a  lattice tilt corresponding to a static potential with uniform gradient. The major theoretical breakthrough is credited to Sachdev, Sengupta, and Girvin \cite{Sach02} who mapped the tilted Bose-Hubbard  model with integer filling factor onto an effective Ising spin system to demonstrate that the MI-state evolves to the density-wave (DW) state as the growing potential gradient traverses the point of quantum phase transition, the DW sate being an ordered particle-hole excitations of the MI state in which empty lattice sites alternate with doubly occupied ones. Later on competing density-wave orders in a one-dimensional hard-boson model were described \cite{Fend04} and theoretical approaches were developed for both quench \cite{Seng04,Rubb11} and adiabatic \cite{Polk05} dynamics across quantum critical points. The  quantum phase transition predicted in Ref.~\cite{Sach02} was confirmed in the pioneering experiment \cite{Simo11} in 2011 and later in a more clear form in the experiment \cite{Mein13},  where a considerable reduction of the residual harmonic confinement was achieved. Importantly, the DW states corresponds to the  maximally possible number of doubly occupied sites (doublons). Thus, by measuring the number of doublons, one can access how close   the final state of the system is to the target DW state \cite{Mein13}.

The mentioned analogy between Ising spin chains and boson systems brought up a new trend in physics of cold atoms \cite{Spie11} and initiated studies on tilted Bose-Hubbard model including doublon production through dielectric breakdown \cite{Ecks13,Oka12}; MI dynamics in parabolic confinement \cite{Lund11}; photon-assisted tunneling for strongly correlated Bose gas \cite{Ma11,Balz14}; impact of quantum quench on Bloch oscillations \cite{64,Mahm14}; upward propagation in the gravity field \cite{ Cai10}; long-range tunneling \cite{Mein14b,Quei12}; and formation of quantum carpets \cite{Muno15}. Spin analogies for various involved configurations of lattice and/or inter-particle interactions were proposed \cite{Piel11,Piel12}. Finally, non-equilibrium dynamics of MI state in relation to  the effective Ising model was considered  \cite{Kolo12a,Kolo12} where the defect density and order parameter correlation function have been  computed. Recent progress in the filed of out-of-equilibrium dynamics in strongly interacting one-dimensional systems is reviewed in \cite{Dale14} while the numerical techniques for solving the Bose-Hubbard model with a tilt are addressed in \cite{Parr14}.

In this paper we approach the problem from the different point of view. Namely, instead of mapping the bosonic system into a spin system, we employ the theory of multi-level Landau-Zener (LZ) tunneling. This theory is an extension of the common Landau-Zener theory from two onto  many (including the case of infinitely many) levels, showing a structured avoided crossing \cite{Pokr01,Ostr06,Sini13,Shyt04,Toma08}.  We identify the diabatic and adiabatic regimes of the multi-level LZ tunneling and derive asymptotic equations for the number of  produced doublons depending on the system parameters. Importantly, our approach admits a straightforward generalization onto  two-dimensional tilted lattices, which have so far attracted less attention.

%%%%%%%%%%%%%%%%%%%%%%%%%%%%%%%%%%%%%%%%%%%%%%%%%%%
\section{The model and main equations}
\label{secA}

First we discuss the one-dimensional case. We consider a unit-filled Bose-Hubbard model with the following Hamiltonian%**********************************************************
\begin{equation}
\label{a1}
\widehat{H}=-\frac{J}{2}\sum_l \left(\hat{a}^\dagger_{l+1}\hat{a}_l + h.c.\right)
+\frac{U}{2} \sum_l \hat{n}_l(\hat{n}_l -1) + F\sum_l l \hat{n}_l \;,
\end{equation}
where $J$ is the hopping matrix element, $U$ the microscopic interaction constant, and the external field $F=F(t)$ is assumed to slowly increase from zero to a value above the interaction constant $U$. Through the paper we shall use the periodic boundary condition, which can be imposed after applying the gauge transformation for the external field. Thus we simulate dynamics of the following system,
%**********************************************************
\begin{equation}
\label{a2}
\widehat{H}(t)=-\frac{J}{2}\sum_{l=1}^L \left(\hat{a}^\dagger_{l+1}\hat{a}_l e^{i\theta(t)}+ h.c.\right)
+\frac{U}{2} \sum_{l=1}^L \hat{n}_l(\hat{n}_l -1) \;,
\end{equation}
where $\theta(t)=\int_0^t F(t'){\rm d}t'$ and $\hat{a}_{L+1}\equiv\hat{a}_l$.  The periodic boundary condition facilitates studying of the thermodynamic limit $N=L\rightarrow\infty$. Going ahead, we mention that convergence of the results towards the thermodynamic limit crucially depends on the sweeping rate $\nu={\rm d} F/{\rm d}t$ which is our control  parameter.  We found rapid convergence for large $\nu$, while it is asymptotically slow if $\nu\rightarrow 0$.

Next we comment on the Hilbert space of the Hamiltoniam (\ref{a2}). For $U\gg J$ and unit filling factor the whole Hilbert space of  $L$ bosons can be truncated to the subspace spanned by the Fock states $|{\bf n}\rangle=|n_1,n_2,\ldots,n_L\rangle$, where the number of atoms in a given site can be zero, one, or two. Accuracy of this approximation is obviously controlled by the ration $J/U$ which we choose to be $<0.1$. The introduced subspace is reduced further by noticing that the periodic  boundary condition conserves the total quasimomentum $\kappa$. Thus, the Hamiltonian matrix can be split into $L$ blocks by introducing the translationally invariant Fock states. We are interested only in $\kappa=0$ block because it contains the initial MI state.  In what follows we shall refer to the specified Hilbert space as the {\em doublon Hilbert space} and denote its dimension by ${\cal N}_D$. Two states of our prime interest in this Hilbert space are the MI state
%**********************************************************
\begin{displaymath}
|MI\rangle=|1,1,1,1,\ldots\rangle \;,
\end{displaymath}
and the DW state
%**********************************************************
\begin{equation}
\label{a5}
|DW \rangle = \frac{1}{\sqrt{2}}(|0,2,0,2,\ldots \rangle + |2,0,2,0,\ldots \rangle) \;,
\end{equation}
where the symmetric form of the DW state is obviously due to the periodic boundary condition. %(In fact, every basis state in the doublon Hilbert space is given by the symmetrized sum of the Fock states which are related to each other by the cyclic permutation.)
%#############################################
\begin{figure}
\center
\includegraphics[width=8cm,clip]{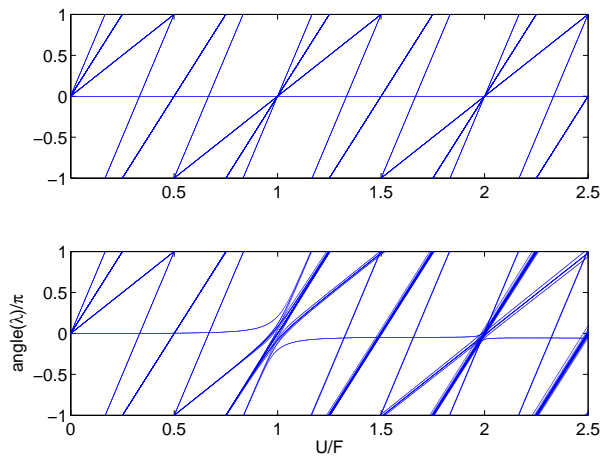}
\caption{Spectrum of the Floquet operator (\ref{a3}) as the function of $U/F$ for $J=0$, upper panel, and $J=0.04U$, lower panel. The system size  $L=6$, where dimension of the doublon Hilbert space ${\cal N}_D=26$.}
\label{fig1}
\end{figure}

Finally we introduce the instantaneous Floquet operator which will be in the core of our analytical approach. To calculate this operator we fix $F$, so that $\theta(t)=Ft$ in Eq.~(\ref{a2}), and calculate the evolution operator over the Bloch period $T=2\pi/F$:
%**********************************************************
\begin{equation}
\label{a3}
\widehat{W}=\widehat{\exp}\left(-i\int_0^{T} \widehat{H}(t) {\rm d} t \right) \;.
\end{equation}
Let us briefly discuss the spectrum of the operator (\ref{a3}). It is convenient to begin with the case of zero hopping where the Fock states $|{\bf n}\rangle$ are also eigenstates of the Floquet operator:
%**********************************************************
\begin{equation}
\label{a4}
\widehat{W}|{\bf n}\rangle=\lambda |{\bf n}\rangle \;,\quad
\lambda=\exp\left(-i\frac{\pi U}{F} \sum_{l=1}^L n_l(n_l-1)\right) \;.
\end{equation}
Plotting eigenphases ${\rm angle}(\lambda)=i\log(\lambda)$ as the function of $1/F$ we obtain a characteristic pattern shown in Fig.~\ref{fig1}(a). Each line in this figure is associated  with a fixed number of doublons: the line with zero slope is the MI state, the first line with nonzero slope is one-doublon states, etc.,  and the line with the maximal slope is the DW state.
%#############################################
\begin{figure}
\center
\includegraphics[width=8cm,clip]{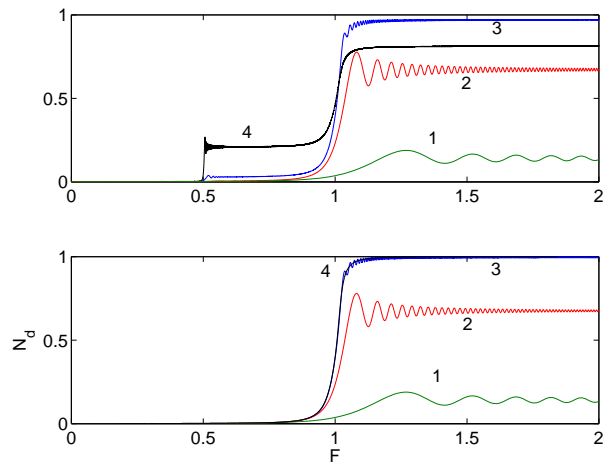}
\caption{The mean number of doublons $N_d$ normalized to $N_{max}=L/2$ as the function of time for different sweeping rates $\nu$,  calculated by using  the doubloon Hilbert space (upper panel) and $j=1$ resonant subspace (lower panel). Parameters are $J=0.02U$, $L=6$, and $\nu=10^{-1}/\pi$ (marked by 1), $\nu=10^{-2}/\pi$ (marked by 2), $\nu=10^{-3}/\pi$ (marked by 3), and $\nu=10^{-4}/\pi$ (marked by 4). The time is related to $F$ as $F=U \nu t$.}
\label{fig2}
\end{figure}

For $J=0$ the majority of levels in Fig.~\ref{fig1}(a) are multiply degenerate, with the MI and DW states being obvious exclusions. Non-zero $J$ removes the degeneracy and originates the multi-level avoided crossings at $F=U/j$ where $j$ is a positive integer number, see Fig.~\ref{fig1}(b). These avoided crossings are associated with the first-order ($j=1$), second-order ($j=2$), etc., resonant tunneling of atoms in the tilted lattice. Our ultimate goal is to calculate the number of doublons $N_d$ as we subsequently traverse the multi-level avoided crossings in Fig.~\ref{fig1}(b) by tilting the lattice from $F=0$ to $F>U$. For the purpose of future discussions Fig.~\ref{fig2}(a) shows $N_d=N_d(t)$ which is obtained by the straightforward numerical simulations of time-evolution of the system (\ref{a2}), where we used the linear ramp for the static field, i.e., $F=\nu t$ (and, hence, $\theta(t)=\nu t^2/2$).  For a large swiping rate $N_d$ is seen to approach zero while for a small $\nu$ it evolves in a step-wise manner, where the positions of the steps correlate with positions of the avoided crossings in Fig.~\ref{fig1}(b).

%%%%%%%%%%%%%%%%%%%%%%%%%%%%%%%%%%%%%%%%%%%%%%%%%
\section{Multi-level Landau-Zener tunneling}
\label{secB}

We shall analyze each multi-level avoided crossing (MLAC) separately. It is instructive to begin with the case $j=1$ which corresponds to the first-order resonant tunneling.

\subsection{The case $j=1$}

\begin{figure}
\center
\includegraphics[width=8cm,clip]{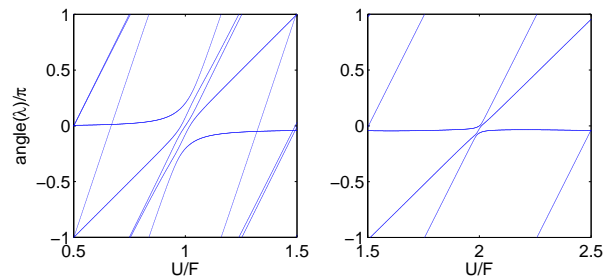}
\caption{MLAC at $F=U$ (left panel) and $F=U/2$ (right panel). Parameters are $L=6$, and $J=0.02U$ in the left panel and $J=0.04U$ in the right panel.}
%Dimension of the resonant subspaces ${\cal N}_R=5$, that should be compared with dimension of the doublon Hilbert space ${\cal N}_D=26$ and dimension of the $\kappa=0$ subspace ${\cal N}=79$ of the total Hilbert space.}
\label{fig3}
\end{figure}

The first step in the analysis is to identified the {\em resonant subspace}. In the case $j=1$ the resonant subspace consists of doublon Fock states with additional constraint that two doublons cannot occupy the nearest sites \cite{Sach02}.  This constraint drastically decreases the dimension of the doublon Hilbert space through removing all irrelevant states, i.e., those that cannot be excited from the initial MI state by means of the first-order resonant tunneling. For the parameters of Fig.~\ref{fig1}(b) the relevant states are shown in Fig.~\ref{fig3}(a). Notice that the MI state (the horizontal line) is analytically connected with the DW state (the line with the maximal slope). An important quantity which can be extracted from the depicted  spectrum is the minimal gap $\Delta$ separating the lowest level, i.e.,  the level which analytically connects the MI and DW states, from the next level. Since  number of levels in MLAC progressively increases with $L$ [see Eq.~(\ref{NR}) in the Appendix A] the gap $\Delta$ tends to zero as $L$ tends to infinity and  we found that with good accuracy
%**********************************************************
\begin{equation}
\label{b0}
\Delta=8J/L \;.
\end{equation}

After truncating the doublon Hilbert space to the resonant subspace the problem can be reformulated as a problem of multi-level Landau-Zener tunneling \cite{Pokr01,Ostr06,Sini13}. This theory deals with systems of the following type,
%**********************************************************
\begin{equation}
\label{b1}
i\frac{{\rm d}\psi}{{\rm d}t}=(H_1+t H_2)\psi \;,\quad  -\infty<t<\infty \;,
\end{equation}
where $H_1$ and $H_2$ are two matrices or two Hamiltonians. For the currently considered  case $j=1$ these  Hamiltonians were found in Refs.~\cite{Sach02,Simo11}, where they were expressed through the spin operators of the effective spin system. In our analysis we do not use this mapping and calculate the matrices $H_1$ and $H_2$ directly from the original Hamiltonian. Given $F=\nu t$ the instantaneous spectrum of the effective Hamiltonian $H(t)=H_1+tH_2$ coincides with the spectrum of the Floquet operator shown in Fig.~\ref{fig3}(a) after folding the former into the fundamental energy interval $-F/2\le E < F/2$.
%#############################################

As soon as we know the matrices $H_1$ and $H_2$ we can use a number of rigorous results from the theory of multi-level LZ tunneling. Let us define the {\em integral} probability of LZ tunneling across MLAC  as
%**********************************************************
\begin{equation}
\label{b2}
{\cal P}_{LZ}=1-\frac{N_d(t=\infty)}{N_{max}} \;,
\end{equation}
where $N_{max}=L/2$ is the maximally possible number of doublons. We mention that definition (\ref{b2}) differs from the standard definition of multi-level LZ tunneling which involves ${\cal N}_R({\cal N}_R+1)/2$ transition probabilities between the instantaneous states of the system. (Here ${\cal N}_R$ is the dimension of the resonant subspace which determines the size of the matrices $H_1$ and $H_2$.) The advantage of Eq.~(\ref{b2}) is that it converges in the thermodynamic limit. This allows us to use terminology of the two-level Landau-Zener problem: we shall call transition across MLAC diabatic if ${\cal P}_{LZ}\approx1$ and adiabatic if ${\cal P}_{LZ}\approx 0$.

\begin{figure}
\center
\includegraphics[width=8cm,clip]{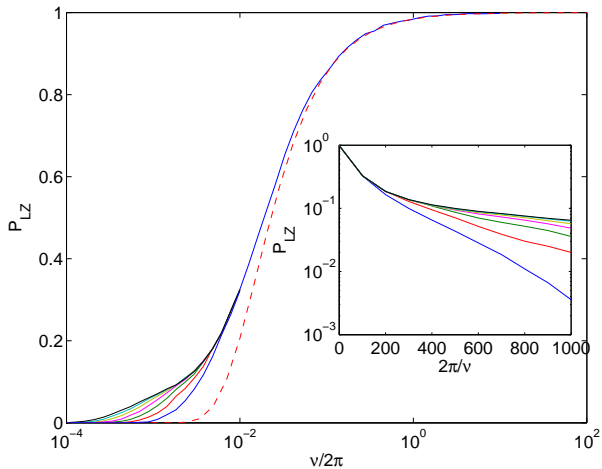}
\caption{Probability of LZ tunneling across $j=1$ MLAC as the function of the sweeping rate $\nu$ for different system size $L=6,8,10,12,14,16,18$. The corresponding dimensions  of the resonant subspaces are ${\cal N}_R=5,8,15,31,64,143, 329$. The hopping matrix element $J=0.02U$. The inset shows the same data in the semi-logarithmic scale as the function of $1/\nu$.}
\label{fig4}
\end{figure}

We begin with the diabatic regime. Using Eq.~(13) in Ref.~\cite{Ostr06} it can be proved that in the limit of large $\nu$ the integral probability is given by
%**************************************************
\begin{equation}
\label{b3}
{\cal P}_{LZ}=\exp\left( -\pi \frac{J^2}{\nu U} \right) \;.
\end{equation}
Here `large $\nu$' means that  $P_{LZ}$ is close to unity. Notice that $P_{LZ}\approx 1$ does not imply occupation of the MI state to be close to unity -- on the contrary, in the thermodynamic limit it goes to zero. Accuracy of Eq.~(\ref{b3}) is illustrated in the main panel in Fig.~\ref{fig4}. In this figure the dashed line is Eq.~(\ref{b3}) and the solid lines are numerical results for different system size $6\le L\le18$.
%#############################################

We proceed with the opposite case of small $\nu$. Here one clearly sees the finite-size effect due to a finite gap $\Delta$ between the lowest level and the next level in Fig.~\ref{fig3}(a). Because of the gap the system sooner or later enters the usual adiabatic regime where the probability to stay  in the lowest level approaches unity while the probability to appear in the next level is an exponentially small value  given by the celebrated Landau-Zener equation: $P\sim\exp\left( -const \Delta^2/\nu \right)$. Provided $L$ is finite, this equation also captures the functional dependence of ${\cal P}_{LZ}$ in the limit $\nu\rightarrow 0$. Namely,
%**********************************************************
\begin{equation}
\label{b4}
{\cal P}_{LZ}=\frac{2}{L}\exp\left( -const\frac{\Delta^2(L)}{\nu} \right) \;.
\end{equation}
This asymptotic behavior is exemplified in the inset in Fig.~\ref{fig4} which shows logarithm of ${\cal P}_{LZ}$ as the function of the inverse swiping rate.

Since the gap $\Delta(L)$ in Eq.~(\ref{b4}) vanishes in the thermodynamic limit, we obtain completely different result if the limits $\nu\rightarrow 0$ and $L\rightarrow\infty$ are exchanged. To find ${\cal P}_{LZ}$ in this case (i.e., to find the limiting curve in Fig.~\ref{fig4}) we use the ansatz
%**************************************************
\begin{equation}
\nonumber
{\cal P}_{LZ}=\exp\left[ f\left(\frac{1}{\nu}\right)\right] \;,
\end{equation}
where $f(\beta)$ is some function of the argument $\beta=1/\nu$. Referring to the inset in Fig.~\ref{fig4}, the derivative of this function  with respect to $\beta$ at the point $\beta=1/\Delta^2$  (here we set $const=1$) takes the value $-\Delta^2$. Thus we have
%**************************************************
\begin{equation}
\nonumber
\frac{{\rm d}f}{{\rm d}\beta}=-\frac{1}{\beta} \;,
\end{equation}
which gives $f=-\ln \beta$ and, hence
%**************************************************
\begin{equation}
\label{b7}
{\cal P}_{LZ} \sim 1/\beta\equiv\nu \;,\quad \nu\rightarrow 0 \;.
\end{equation}
The obtained estimate is in qualitative agreement with numerical results of Ref.~\cite{Kolo12} where ${\cal P}_{LZ}$ was argued to scales as ${\cal P}_{LZ}\sim \nu^{1/2}$ in the thermodynamic limit.

%%%%%%%%%%%%%%%%%%%%%%%%%%%%%%%%%%%%%%%%%%%%%%
\subsection{The case $j=2$}

In the case $j=2$, which corresponds to the second-order resonant tunneling, the spectrum of the Floquet operator in the resonant subspace is depicted in Fig.~\ref{fig3}(b). For the considered $L=6$ the resonant subspace consists of three Fock states: the MI state $|111111\rangle$, one-doublon state  $|012111\rangle$, which is resonantly related to the MI state through the intermediate state   $|021111\rangle$, and two-doublon state  $|012012\rangle$, which is related to one-duoblon state through the state $|012021\rangle$. (It is implicitly assumed that all these states are symmetrized by using cyclic permutation to satisfy the conservation law for the total  quasimomentum.) To find MLAC shown in  Fig.~\ref{fig3}(b) one first calculates the Floquet operator keeping the intermediate states and then eliminates them by projecting this operator onto the basis of the resonant states. This results in the effective Hamiltonian where the resonant states are directly related to each other by the transition matrix elements which are proportional to $J^2/U$. Thus we can use the results of the previous subsection with some minor modifications. Firstly, the maximally possible doublon number $N_{max}=L/3$ but not $L/2$. Secondly, the critical value of the swiping rate $\nu$ which separates the dibasic and adiabatic regimes of the multi-level LZ tunneling scales as $J^4$ but not $J^2$.

%%%%%%%%%%%%%%%%%%%%%%%%%%%%%%%%%%%%%%%%%%%%%%
\section{Dynamics of doublon number}
\label{secC}

In the previous section we considered different regimes of LZ tunneling across a MLAC. It was argued, in particular, that the adiabatic regime is sensitive to the  system size. This result, however, is more of academic than of practical interest. In fact, in the laboratory experiment one deals with an ensemble of 1D lattices where the lattice lengths are determined by the distances between defects in the initial MI state. Thus, the system size $L$ is, strictly speaking,  not known. At the same time, as it is seen in Fig.~\ref{fig4} an error in the dependence ${\cal P}_{LZ}={\cal P}_{LZ}(\nu)$ due to unknown $L$ never exceeds few percents. This allows us to make reliable predictions  by analyzing the lattices of a rather small size. With this in mind we address dependence $N_d=N_d(t)$ in the limit of small $\nu$, which is of prime experimental interest.

Let us assume the sweeping rate $\nu$ to be small enough to ensure truly adiabatic regime. In the other words, we follow the lowest level in Fig.~\ref{fig3}(a) which analytically connect the MI state with the DW state. Denoting by $|\Psi(F)\rangle$ the instantaneous eigenstate of the Floquet operator associated with this level we have
%********************************************
\begin{eqnarray}
\label{c1}
N_d(F)=\langle \Psi(F) |\hat{D}|\Psi(F)\rangle \;,
\end{eqnarray}
where $\hat{D}$ is the doublon number operator. Below we display analytical solutions of Eq.~(\ref{c1}) for $L=2$ and $L=4$ and compare them with the numerical solutions for $L\rightarrow\infty$. It should be mentioned that Eq.~(\ref{c1}) rapidly converges as $L$ is increased and the corresponding curves become undistinguishable in the linear scale if $L\ge 8$.

For $L=2$ the dimension of the resonant subspace ${\cal N}_R=2$ and the problem reduces to diagonalization of $2\times2$ matrix
%************************************************************************
\begin{displaymath}
H(F)=\left(
\begin{array}{cc}
 0 & -J \\
-J & \delta(F) \\
\end{array}
\right) \;,
\end{displaymath}
where $\delta(F)=U-F$. For the mean number of doublons this model gives
%***********************************************
\begin{eqnarray}
\label{c3}
\frac{N_d(F)}{N_{max}}=\frac{{\left(\delta-\sqrt{\delta^2+4J^2} \right)}^2}
{4J^2+{\left(\delta-\sqrt{\delta^2+4J^2} \right)}^2} \;.
\end{eqnarray}
%
%The solution (\ref{c3}) is depicted in Fig.~\ref{fig6}(a) by the dotted line.
Next consider $L=4$. In this case ${\cal N}_R=3$ and
%***********************************************************
\begin{displaymath}
H(F)= \left(
\begin{array}{ccc}
0 & -\sqrt{2}J & 0 \\
-\sqrt{2}J & \delta(F) & -J \\
0& -J& 2\delta(F)
\end{array}
\right) \;.
\end{displaymath}
After some algebra we get
%****************************************************
\begin{eqnarray}
\label{c4}
\frac{N_d(F)}{N_{max}}=  %\frac{1}{2}\cdot
\frac{(E^2-\delta^2)^2+2J^2(E+\delta)^2}
{2J^2(E-\delta)^2+J^2(E+\delta)^2+(E^2-\delta^2)^2} \;,
\end{eqnarray}
where $E=E(F)$ denotes the position of the lowest level:
%*********************************************
\begin{displaymath}
E(F)=2\sqrt{\frac{3J^2+\delta^2}{3}}\cos\left(\frac{\eta+2\pi}{3}\right) \;,
\end{displaymath}
%
%****************************************************
\begin{eqnarray}
\nonumber
\eta(F)=\pi\theta(\delta)-
\arctan\left(\frac{2\sqrt{{\left(\frac{3J^2+\delta^2}{3}\right)}^3-{\left(\frac{J^2\delta}{2}\right)}^2}}{J^2\delta}\right) \;.
\end{eqnarray}
We found that there is no need to consider the next approximation because Eq.~(\ref{c4}) reproduces the results for $L\rightarrow\infty$ with accuracy higher than one persent. Thus, for practical purpose one can use Eq.~(\ref{c4}) or even simpler Eq.~(\ref{c3}). It follows from these equations that the characteristic width of the step for $N_d(F)$ is proportional to $J$.

\begin{figure}
\center
\includegraphics[width=8cm,clip]{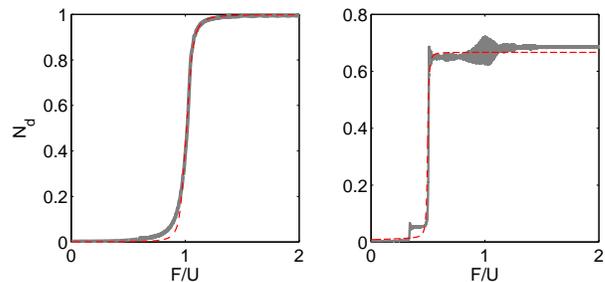}
\caption{Number of doublons as the function of $F=F(t)$ for two different protocols: a piece-wise ramp with  $\nu=1.25\cdot10^{-2}$  in the interval $0\le F/U<0.6$ and  $\nu=2.5\cdot10^{-4}$ in the interval $0.6\le F/U<2$ (left panel), an the linear ramp in the interval $0\le F/U<1$ with the rate $\nu=2\cdot 10^{-5}$ (right panel). The dashed lines are analytical results Eq.~(\ref{c4}) and Eq.~(\ref{NdF}), respectively. The system size $L=8$, where dimension of the doublon Hilbert space ${\cal N}_D=142$. The hopping matrix element $J=0.04$.}
\label{fig6}
\end{figure}

Similar equation can be derived for the second-order resonant tunneling at $F=U/2$, see Eq.~(\ref{NdF}) in the Appendex B.  The dependences (\ref{c4}) and (\ref{NdF}) are shown in Fig.~\ref{fig6} by the dashed lines. It is interesting to compare Eq.~(\ref{c4}) and Eq.~(\ref{NdF}) against direct numerical simulations of the doublon dynamics, see solid lines in Fig.~\ref{fig6}.  We mention that in these simulations we use the doublon Hilbert space and, hence, no resonant approximations are involved. In Fig.~\ref{fig6}(b) we tilt the lattice to $F=U$ by using the linear ramp with the rate $\nu=2\cdot 10^{-5}$, which is small enough to ensure the adibatic regime for MLAC at $F=U/2$. In Fig.~\ref{fig6}(a) we tilt the lattice to $F=2U$ and use a protocol with two different rates: in the interval $0\le F/U<0.6$ the rate  $\nu=1.25\cdot10^{-2}$, which ensures diabatic regime for MLAC at $F=U/2$; in the interval $0.6\le F/U<2$ the rate is changed to $\nu=2.5\cdot10^{-4}$, which insures the adiabatic regime for the second avoided crossing at $F=U$. A good agreement with analytical results is noticed.
%#############################################

%%%%%%%%%%%%%%%%%%%%%%%%%%%%%%%%%%%%%%%%%%%%%%%%%%%%%%
%%%%%%%%%%%%%%%%%%%%%%%%%%%%%%%%%%%%%%%%%%%%%%%%%%%%%%
\section{Two-dimensional lattices}
\label{secD}

In this section we generalize the results of the previous sections onto two-dimensional case,
%********************************************
\begin{eqnarray}
\nonumber
\widehat{H}= -\frac{J_x}{2}\sum_{l,m}  \left(\hat{a}^\dagger_{l+1,m}\hat{a}_{l,m} + h.c.\right) \\
\nonumber
-\frac{J_y}{2}\sum_{l,m}  \left(\hat{a}^\dagger_{l,m}\hat{a}_{l,m+1} + h.c.\right) \\
\nonumber
+\frac{U}{2}\sum_{l,m} \hat{n}_{l,m}(\hat{n}_{l,m}-1)\\
\label{d1}
- F(t)\sum_{l,m}[l \cos\phi + m \sin\phi] \hat{n}_{l,m} \;,
\end{eqnarray}
where, as before, $F(t)$ changes linearly in time with the rate $\nu$, and the initial state of the system is a Mott insulator with unit filling. Like for 1D lattices we shall use the periodic boundary conditions, which are imposed after applying the gauge transformation. Thus we simulate dynamics of finite system of the size $L_x\times L_y$ with the Hamiltonian
%********************************************
\begin{eqnarray}
\nonumber
\widehat{H}(t)= -\frac{J_x}{2}\sum_{l=1}^{L_x} \sum_{m=1}^{L_y}  \left(\hat{a}^\dagger_{l+1,m}\hat{a}_{l,m}e^{-i\theta_x(t)} + h.c.\right) \\
\nonumber
-\frac{J_y}{2}  \sum_{l=1}^{L_x} \sum_{m=1}^{L_y} \left(\hat{a}^\dagger_{l,m+1}\hat{a}_{l,m}e^{-i\theta_y(t)} + h.c.\right) \\
\label{d2}
+\frac{U}{2}\sum_{l=1}^{L_x} \sum_{m=1}^{L_y} \hat{n}_{l,m}(\hat{n}_{l,m}-1)   \;,
\end{eqnarray}
where $\theta_x(t)=\int F_x(t) {\rm d}t$ and  $\theta_y(t)=\int F_y(t) {\rm d}t$. The main difference and challenge of the 2D system (\ref{d2}) as compared to the 1D system (\ref{a2}) is sensitivity to the field orientation. The cases where ${\bf F}$ is exactly aligned  or slightly misaligned with one of the primary axes of the lattice have been analyzed in the recent work \cite{preprint}. Here we address another important case where ${\bf F}$ is {\em strongly} misaligned with the primary axes. It will be shown below that strongly misaligned 2D lattices are closer to the one-dimensional situation than the aligned lattices.

%%%%%%%%%%%%%%%%%%%%%%%%%%%%%%%%%%%%%%%%%%%%%%%%%
\subsection{Floquet operator}

\begin{figure}
\center
\includegraphics[width=8cm,clip]{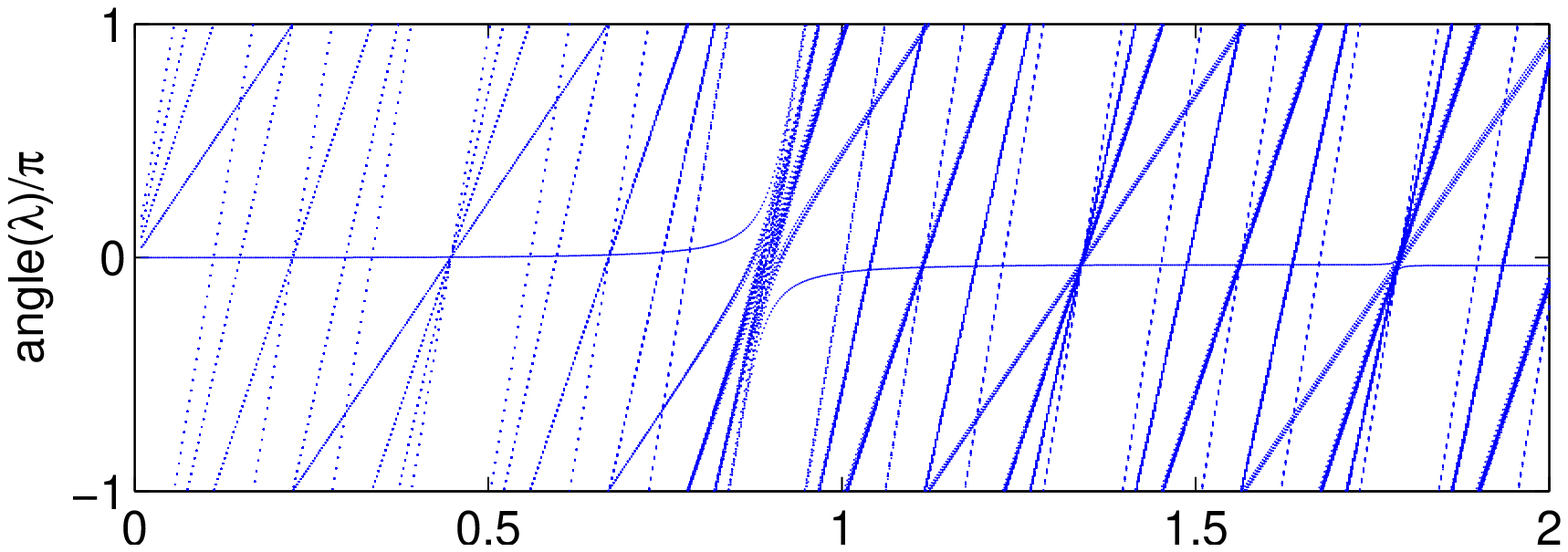}
\includegraphics[width=8cm,clip]{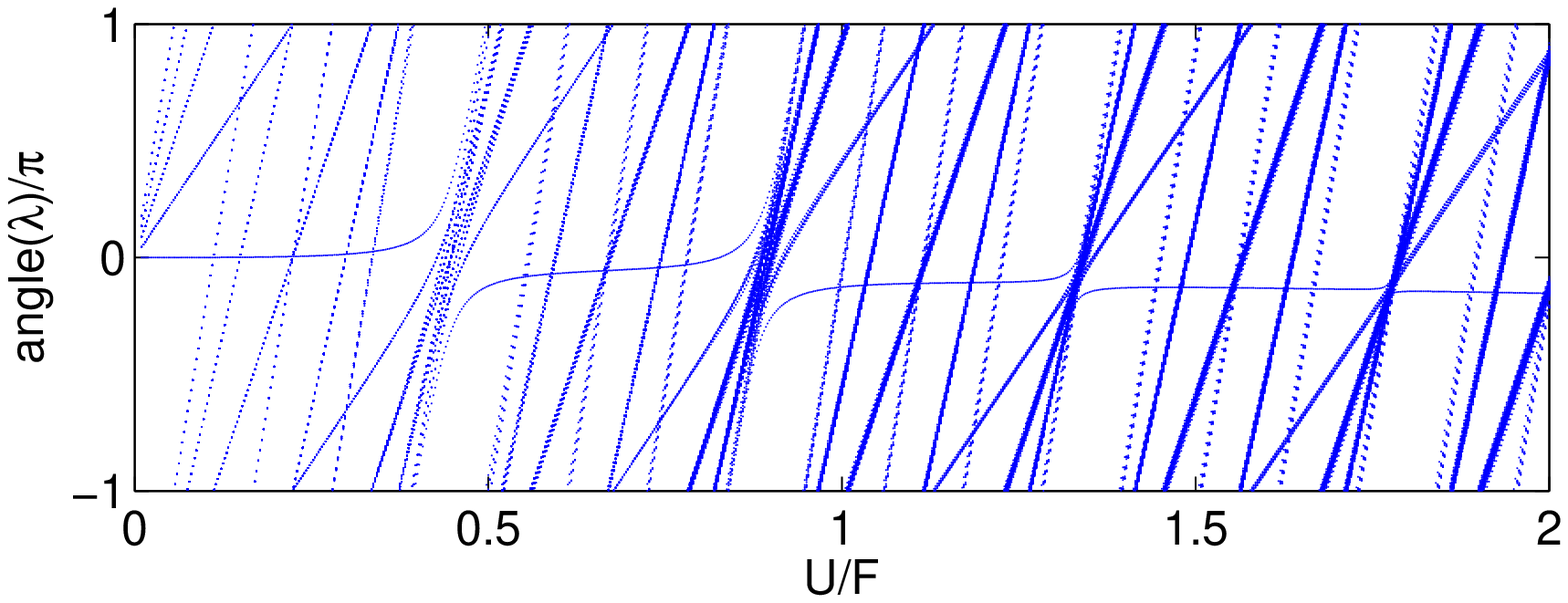}
\caption{Eigenphases of the Floquet operator as the function of $1/F$ for $F_x/F_y=1/2$,  and $J_x=0$, upper panel, and $J_x=0.04U$, lower panel. The lattice size is $2\times4$ (dimension of the doublon Hilbert space ${\cal N}_D=147$), the hopping matrix element $J_y=0.02U$.}
\label{gig1}
\end{figure}

To be specific we shall consider the field orientation $F_x/F_y\approx 1/2$ and we begin with the case where $F_x/F_y=1/2$ exactly. In this case we can introduce the Floquet operator,
%*************************************************
\begin{equation}
\label{d3}
\widehat{W}=\widehat{\exp}\left(-i \int_0^T \widehat{H}(t) {\rm d}t\right) \;,
\end{equation}
where $T=2\pi/F_x=4\pi/F_y$ is the common Bloch period. Similar to 1D tilted lattices we restrict ourselves by the doublon Hilbert space where $n_{l,m}\le2$. The validity of this, not obvious for 2D tilted lattices approximation will be checked later on. Using the doublon Hilbert space we calculate the spectrum of the operator (\ref{d3}) and decompose it into $L_x\times L_y$ independent spectra according to the total quasimomentum. As before, we are bound  with the zero quasimomentum subspace because the MI state belongs to this subspace.  The obtained spectrum is shown in Fig.~\ref{gig1}  for  the lattice $2\times4$, $J_y=0.02U$, and $J_x=0$, upper panel, and $J_x=0.04U$, lower panel.
%##########################################

Let us discuss the depicted spectra in more detail. The spectrum in Fig.~\ref{gig1}(a) obviously reproduces the spectrum of two independent 1D lattices of the length $L=L_y$, where MLAC at $F=\sqrt{5}/2$ corresponds to the first-oder tunneling in the $y$ direction.  The spectrum in Fig.~\ref{gig1}(b) contains extra MLAC  at $F=\sqrt{5}$, which corresponds to the first-order tunneling in the $x$ direction, and a number of less pronounced crossings corresponding to the second-order tunneling. In what follows we shall focus on the first-order resonance at  $F=\sqrt{5}/2$.

\begin{figure}
\center
\includegraphics[width=8cm,clip]{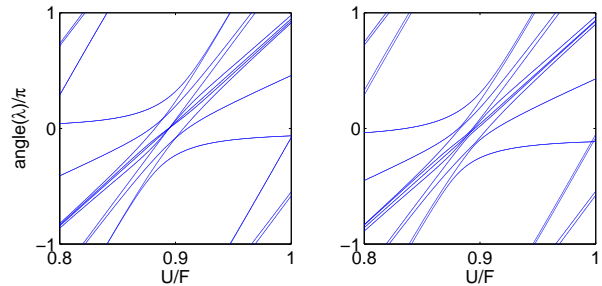}
\caption{The spectrum of the Floquet operator on the resonant subspace (${\cal N}_R=10$) for $J_x=0$, left panel, and $J_x=0.04U$, right panel.}
\label{gig2}
\end{figure}

If $J_x=0$ the doublon Hilbert space can be truncated to the resonant subspace, which is given by the tensor product of two (in general case, $L_x$) 1D resonant subspaces introduced earlier in Sec.~\ref{secB}. The spectrum of the operator (\ref{d3}) on this subspace is shown in Fig.~\ref{gig2}(a). Our particular interest in Fig.~\ref{gig2}(a) is the `lowest' level. Using the fact that two 1D lattices are independent it is easy to prove that this level analytically connects the MI state with the state
%*************************************************
\begin{equation}
\label{d4}
|\psi\rangle=\frac{1}{\sqrt{2}}\left(|DW\rangle +  |\widetilde{DW}\rangle \right) \;,
\end{equation}
where $|DW\rangle$ is the `correlated'  DW state,
%*************************************************
\begin{equation}
\label{d5}
|DW\rangle =\frac{1}{\sqrt{2}}\left[\left(
\begin{array}{cc}
22\\00\\22\\00
\end{array}\right)
+\left(
\begin{array}{cc}
00\\22\\00\\22
\end{array}\right)\right] \;,
\end{equation}
and $|\widetilde{DW}\rangle$ is the  `uncorrelated' DW state,
%*************************************************
\begin{equation}
\label{d6}
|\widetilde{DW}\rangle =\frac{1}{\sqrt{2}}\left[\left(
\begin{array}{cc}
20\\02\\20\\02
\end{array}\right)
+\left(
\begin{array}{cc}
02\\20\\02\\20
\end{array}\right)\right] \;.
\end{equation}
%
%##########################################

Let now $J_x\ne0$ and, hence, two 1D lattices are no more independent.  To treat this situation we shall use the specific perturbation theory over the parameter $J_x$. The procedure involves two steps and goes as follows. First we introduce the new basis which diagonalizes  the Floquet operator (\ref{d3}) for $J_y=0$. We shall refer to this basis as  many-body Wannier-Stark  states. If $J_x\ll F_x$ and $F_x\ne U$ (the latter condition ensures that there is no resonant tunneling in the $x$ direction) these many-body Wannier-Stark states can be approximated by the Fock states which, however,  have slightly different energies
%*************************************************
\begin{equation}
\label{d7}
E_i=E^{(0)}_i+\Delta E_i \;, \quad E^{(0)}_i=\langle{\bf n}_i|  \frac{U}{2}\sum_{l,m} \hat{n}_{l,m}(\hat{n}_{l,m}-1) |{\bf n}_i\rangle \;.
\end{equation}
We find the energies $E_i$ by calculating diagonal elements of the Floquet operator for $J_y=0$., i.e., by dropping the second term in the Hamiltonian (\ref{d2}). Notice that for $J_y=0$ the system becomes quasi one-dimensional. For this reason the above introduced  correction $\Delta E_i$ to the energy of $i$th Fock states can be found semi-analytically by using simple combinanatorics.
%is given by the sum of $L_y$ terms, where each term is uniquely determined by the rows of this Fock state. Thus one can find corrections $\Delta E_i$ also by means of combinatorics. For example, for the lattice $2\times4$ the corrections $\Delta E_i$ can take one of six values: $4b$, $2b$, $b/2$, $0$, $-b$, $-2b$, where $b$ is proportional to $J_x^2$ and smoothly depends on $F_x$, see Fig.~\ref{gig6}.
%%##########################################
%\begin{figure}
%\center
%\includegraphics[width=8cm,clip]{gig6.eps}
%\caption{Corrections $\Delta E_i$ to the energies of the resonant Fock state for $L_x=2$, left panel, and $L_x=4$ right panel.}
%\label{gig6}
%\end{figure}

\begin{figure}
\center
\includegraphics[width=8cm,clip]{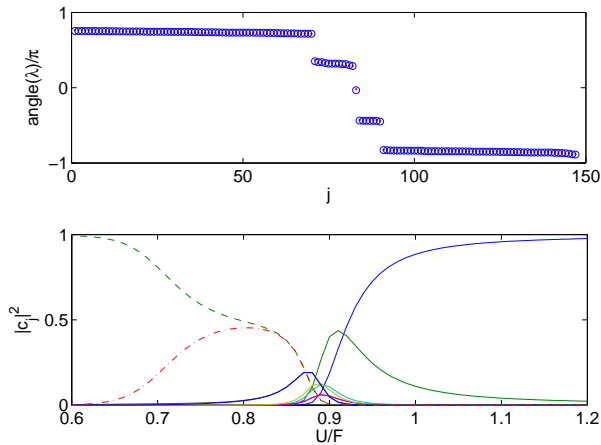}
\caption{Upper panel: Exact (dots) and approximate (open circles)  eigenphases of the Floquet operator for $F_x/F_y=1/2$. The other parameters are $U/F=0.8$, $J_y=0.02U$, and $J_x=0.04U$. Lower panel: Squared modulus of the expansion coefficients $c_i$ for the eigenstate associated with the lowest level in Fig.~\ref{gig2}(b). The coefficients in front of the correlated and uncorrelated DW states are marked by the dashed and dash-dotted lines, respectively. The hopping matrix element $J_x$ is set to a very small yet finite value $J_x=0.001U$. }
\label{gig3}
\end{figure}

In the second step we calculate the Floquet operator (\ref{d3}) approximately, by dropping the first term in the Hamiltonian (\ref{d2}) and simultaneously correcting the energies of the Fock states.  This again reduces the 2D problem to a quasi 1D problem,  where the $x$ degree of freedom is now taken into account by non-zero $\Delta E_i$. The accuracy of the method is illustrated in Fig.~\ref{gig3}(a) which compares the eigenphases of the exact and approximate Floquet operators for $J_y=0.02U$ and $J_x=0.04U$.
%##########################################

The described approach, although perturbative, has several advantages over the straightforward diagonalization of the Floquet operator. Firstly, it allows us to treat lager lattices by reducing the 2D problem to the sequence of two quasi 1D problems. Secondly, it can be also used in the case of irrational orientations of the field, where one has no common Bloch period. Finally,  it justifies resonant approximation for $J_x\ne0$ and provides a physical interpretation of the numerical results in terms of the energies $\Delta E_i$. The right panel in Fig.~\ref{gig2} shows the spectrum of the Floquet operator for $J_x=0.04U$, calculated by using the resonant Hilbert space. It is seen that the lowest level is now separated from the next level by a finite gap $\widetilde{\Delta}$. The size of the gap is given by the difference  between the energy corrections $\Delta E_i$ to the correlated DW state (\ref{d5}) and uncorrelated DW state (\ref{d6}), which was found to scale as
%*************************************************
\begin{equation}
\label{d8}
\widetilde{\Delta} \approx 5J_x^2/U  \;.
\end{equation}
The presence of the gap also breaks the symmetry of the $J_x=0$ problem so that the MI state is now analytically connected with the correlated DW state but not with the symmetric state (\ref{d4}). This is illustrated  in the lower panel in Fig.~\ref{gig3}, which shows expansion coefficients over the Fock basis for the eigenstate $|\Psi(F)\rangle$ associated with the lowest level,
%*************************************************
\begin{equation}
\label{d9}
|\Psi(F)\rangle=\sum_{j=1}^{{\cal N}_R} c_i(F) |{\bf n}_i\rangle  \;.
\end{equation}
It is seen in Fig.~\ref{gig3} that, as soon as $J_x\ne0$, the coefficient in front of the uncorrelated DW state tends to zero while the coefficient in front of the correlated DW state tends to unity. This result holds for arbitrary $L_x$ where we have several uncorrelated DW states. For example, for $L_x=4$ these are
%*************************************************
\begin{equation}
\label{d10}
\left(
\begin{array}{cccc}
2220\\0002\\2220\\0002
\end{array}\right)
\;,\quad
\left(
\begin{array}{cccc}
2200\\0022\\2200\\0022
\end{array}\right)
\;,\quad
\left(
\begin{array}{cccc}
2020\\0202\\2020\\0202
\end{array}\right) \;.
\end{equation}
(Unlike Eq.~(\ref{d5}) and Eq.~(\ref{d6}) here we display not  symmetrized  Fock states --  the symmetrization procedure is assumed implicitly.) We found that the closest to the energy of the correlated DW state is the uncorrelated DW state which is obtained from the former by shifting one column, like the first state in Eq.~(\ref{d10}). Furthermore,  the energy difference between these two states (i.e, the difference between associated corrections $\Delta E_i$) is essentially independent of $L_x$. Thus, the correlated DW state is separated from a bundle of uncorrelated DW states by a finite gap,  where Eq.~(\ref{d8}) provides an estimate for the gap size.

%%%%%%%%%%%%%%%%%%%%%%%%%%%%%%%%%
\subsection{Dynamics of doublon number}

\begin{figure}
\center
\includegraphics[width=8cm,clip]{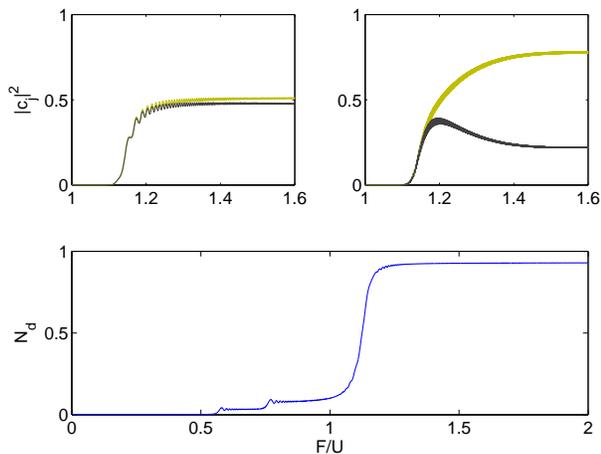}
\caption{Lower panel: The mean number of doublons $N_d$ normalized to $N_{max} = L_xL_y/2$ as the function o$F=\nu t$. Parameters are $J_y = 0.02U$, $J_x=0.005U$, $L_x=2$, $L_y=4$, and $\nu=10^{-3}/2\pi$. Upper panels shows occupations of the correlated and uncorrelated DW states for the rates $\nu=10^{-3}/2\pi$ and $\nu=10^{-4}/2\pi$.}
\label{gig5}
\end{figure}

This subsection presents numerical solutions of the time-dependent Schr\"odingier equation with the Hamiltonian (\ref{d2}) where $F(t)=\nu t$.  Simulations are performed in the doublon Hilbert space. The lower panel in Fig.~\ref{gig5}(c) shows doublon number $N_d$ as the function of time for the lattice $2\times4$ and $\nu=10^{-3}/2\pi$. As expected, one finds many similarities with  Fig.~\ref{fig2}(a) showing the result for 1D lattices. In particular, small steps are due to the second-order tunneling and the large step at $F=\sqrt{5}/2$ is due to the first-order tunneling.  By using an appropriate protocol for the swiping rate $\nu$ we can ensure the diabatic regime for MLACs associated with the second-order tunneling. Then the main step will be described by Eq.~(\ref{c4}) and $N_d(t)$ approaches $N_{max}=L_xL_y/2$.

The mean number of doublons, however, does not provide the whole information about the final state of the system -- it can be only stated that populations of the correlated and uncorrelated DW states sum up to unity. For this reason we specifically address populations of the states Eq.~(\ref{d5}) and Eq.~(\ref{d6}), see  the upper-left panel in Fig.~\ref{gig5}.  It appears that the chosen rate $\nu$ does not ensure a fully adiabatic regime, so that the populations of the correlated and uncorrelated DW states become almost equal due to LZ tunneling between two lowest levels in Fig.~\ref{gig3}(b). To obtain the correlated DW state as the final state the sweeping rate should be essentially smaller, smaller than the inverse gap (\ref{d10}).  In fact, for $\nu=10^{-4}/2\pi$ we already observe a misbalance in the population [see the upper-right panel in Fig.~\ref{gig5}], which tends to unity for smaller $\nu$ or, alternatively, larger $J_x$.
%##########################################

To conclude this section we comment on truncation of the Hilbert space to the doublon subspace. Validity of this approximation assumes negligible population of the Fock states which may lead to triple occupations of the lattice sites.  We have checked that during adiabatic passage the population of these states is orders of magnitude smaller than the population of the resonant states. As the final check we repeated calculations shown in Fig.~\ref{gig5} by using the whole Hilbert space  -- the results appear to be almost identical. We stress, however, that the truncation of the whole Hilbert space to the doublon Hilbert space and further to the resonant subspace is justified only in the considered case of strong misalignment.  If $\phi\approx0$ we do observe qualitative difference in the doublon dynamics when we truncate the Hilbert space.

%%%%%%%%%%%%%%%%%%%%%%%%%%%%%%%%
\section{Conclusions}

In the work we analyzed the evolution of the Mott-insulator state of cold atoms  in 1D and 2D optical lattice as the lattice is tilted by applying a monotonically increasing static field  $F=F(t)$. The analysis was performed  by using the theory of multi-level Landau-Zener tunneling, properly adopted for the considered problem.

As concerns 1D lattices, the central result of the paper are  Eq.~(\ref{b3}),   Eq.~(\ref{b4}), and  Eq.~(\ref{b7}), which give the number of produced doublons $N_d$ as the function of the swiping rate $\nu={\rm d}F/{\rm d} t$.  We payed particular attention to the adiabatic regime $\nu\rightarrow 0$, where the Mott-insulator state evolves into the density-wave state (empty lattice sites alternating with doublons). For this case we derived analytical expressions which capture the dynamics of the doublon number. It is shown that, having the goal to produce the density wave state, one should use a protocol $F=F(t)$ which ensures diabatic transition of the multi-level avoided crossing at $F=U/2$, which is associated with the second-order tunneling, and adiabatic transition of the multi-level avoided crossing at $F=U$, associated with the first-order tunneling.

The above results can equally be applied to the 2D square lattice, provided that the static field ${\bf F}$ is strongly mismatched with the primary axes of the lattice (for example, $1/3<F_x/F_y<1/2$). In this case the 2D lattices can be viewed as an array of weakly coupled 1D lattices. Correspondently, there are two adiabatic conditions for the rate $\nu$. The first one is deduced from  Eq.~(\ref{b4}). It ensures that every column of the 2D lattice is one-dimensional  density-wave state. The second one requires $\nu \ll 1/\widetilde{\Delta}$ where $\widetilde{\Delta}$ is given in Eq.~(\ref{d8}). It ensures that the column density waves are correlated, i.e., we have empty rows alternating with rows where every site has double occupancy.  %The obtained Eq.~(\ref{d8}) is another central result of the paper.

It might be thought that the field orientation $\phi=\arctan(F_x/F_y)\approx 0$ is more suitable for producing the density-wave state in the square 2D lattice. This, however, is not the case. As shown in Ref.~\cite{preprint}, for $\phi\approx0$ the system has an intrinsic instability due to  high mobility of the quasi-particles (doublons and holes) in the transverse $x$ direction. For a strong misalignment this mobility is suppressed by the Wannier-Stark localization and the quasi-particles are essentially localized in the sites where they were created. Yet, slightly larger than unity localization length introduces non-zero correlations in the $x$ direction, which make it possible to produce the 2D density-wave state from the initial Mott-insulator state by means of the adiabatic passage.

{\em Acknowledgements.}  The author acknowledge financial support from Russian Foundation for Basic Research through the Project No. 16-42-240746. AK acknowledges fruitful discussions with F. Meinert and H.-Ch. N\"agerl.

%%%%%%%%%%%%%%%%%%%%%%%%%%%%%%%%%%%%%%%%%%%%%%%%%%%%%%%

%\bibitem{Mark12}
%M. J. Mark, E. Haller, K. Lauber, J. G. Danzl, A. Janisch, H. P. B\"uchler, A. J. Daley, and H.-C. Na\"gerl,
%{\em Preparation and spectroscopy of a metastable Mott-Insulator state with attractive interactions},
%Phys. Rev. Lett. {\bf 108}, 215302 (2012).

%%%%%%%%%%%%%%%%%%%%%%%%%%%%%%%%%%

\section{Appendix A}

In this appendix we display explicit formulas for the dimension of the Hilbert spaces. The total dimension of the Hilbert space of
the Bose-Hubbard model is given by
the well-known equation
%*******************************************************
\begin{equation}
{\cal N}=\frac{(N+L-1)!}{(N-1)!L!} \;,
\end{equation}
where one should set $N=L$ in the case of unit filling. In the main text we refer to subspace of the
total Hilbert space, which is defined by the condition $n_l\le 2$, as the doublon Hilbert space. It has dimension
%**************************************************
\begin{equation}
{\cal N}_D=\sum_{n=0}^{L/2}\frac{L!}{(L-2n)!(n!)^2} \;.
\end{equation}
Finally, the dimension the $j=1$ resonant subspace is dependent on boundary conditions.
For the closed (Dirichlet) boundaries we have
%*************************************************
\begin{equation}
{\cal N}'_R=\sum_{n=0}^{L/2}\frac{(L-n)!}{(L-2n)!n!} \;,
\end{equation}
while for the periodic boundary condition
%*************************************************
\begin{equation}
\label{NR}
{\cal N}_R= {\cal N}'_R(L)+{\cal N}'_R(L-2) \;.
\end{equation}
Needless to say that ${\cal N}_R<{\cal N}_D<{\cal N}$. For example, for $L=8$ the inequality relation reads as $47<1107<6435$ and
for $L=16$ as $2207<5196627<300540195$. We also mention that in the case of periodic boundary condition dimension of every Hilbert
space can be reduced by factor $L$ if we take into account the conservation of the total quasimomentum.

\section{Appendix B}

To obtain a quantitative description of the second order transition we consider the three-site Bose-Hubbard chain. Projecting Eq.~ (\ref{a2}) onto the basis vectors $|\psi_1\rangle=|1,1,1\rangle$, $|\psi_2\rangle=\frac{1}{\sqrt{3}}\sum_{l=0}^{2} \widehat{\mathcal{T}}^l |0,2,1\rangle$,  and $|\psi_3\rangle=\frac{1}{\sqrt{3}}\sum_{l=0}^{2} \widehat{\mathcal{T}}^l |0,1,2\rangle$ (here $\widehat{\mathcal{T}}$ is the cyclic permutation operator) and removing time dependance from the kinetic term through substitutions
%*****************************************************
\begin{eqnarray}
|\psi_1\rangle=e^{-i(Ut/2-\theta(t))}|\phi_1\rangle \notag \\ %\nonumber
|\psi_2\rangle=e^{-iUt/2}|\phi_2\rangle \notag \\ %\nonumber
|\psi_3\rangle=e^{-i(Ut/2+\theta(t))}|\phi_3\rangle
\end{eqnarray}
we obtain
%*****************************************************
\begin{equation}
H(F)= \left(
\begin{array}{ccc}
-\mu(F) & -\frac{\sqrt{6}J}{2} & 0 \\
-\frac{\sqrt{6}J}{2} & \frac{U}{2} & -\frac{3J}{2} \\
0& -\frac{3J}{2}& \mu(F)
\end{array}
\right)
\end{equation}
where $\mu(F)=\frac{U}{2}-F$. The eigenvalues of this matrix could be found exactly by Cardano's formula. It is much simpler though to find approximate solution for $\mu(F)\approx 0$ because we only interested in the part of the spectrum underlying the second-order resonant transition. After some algebra we have
%********************************************************
\begin{align}
E_{1,2}=\frac{3J^2}{2U^2}\mu-\frac{15J^2}{4U} \notag \\
\mp \sqrt{\left(\frac{3J^2}{2U^2}\mu-\frac{15J^2}{4U}\right)^2+\left(\mu^2-\frac{3J^2}{2U}\mu\right)} + \mathcal{O}\left(\frac{J^2}{U^2}\right), \notag \\
E_3=\frac{U}{2} + \mathcal{O}\left(\frac{J^2}{U^2}\right) \;.
\end{align}
The eigenvalues $E_{1,2}$,  which show an avoided crossing at $\mu(F)\approx 0$, define the effective Hamiltonian. Basing on this effective Hamiltonian the mean number of doublons is given by
%*************************************************************
\begin{equation}
\label{NdF}
\frac{N_d(F)}{N_{max}}\approx \frac{2}{3}\cdot\frac{\left((\mu-E_1)^2 +\frac{9J^2}{4}\right)(\mu+E_1)^2}
{(\mu^2-E^2_1)^2+\frac{9J^2}{4}(\mu+E_1)^2+\frac{3J^2}{2}(\mu-E_1)^2} \;.
\end{equation}

\end{document}